\documentclass[11pt]{article}
\begin{document}
\title{On the AdS/CFT Correspondence and Logarithmic Operators}
\author{S. Moghimi-Araghi \footnote{e-mail: samanimi@rose.ipm.ac.ir}
, S. Rouhani \footnote{e-mail: rouhani@karun.ipm.ac.ir} and M. Saadat
\footnote{e-mail: saadat@mehr.sharif.ac.ir}\\ \\ Department of
Physics, Sharif University of Technology,\\ Tehran, P.O.Box:
11365-9161, Iran\\ Institute for studies in Theoretical physics
and Mathematics,\\ Tehran, P.O.Box: 19395-5531, Iran}
\maketitle

\begin{abstract}
Logarithmic conformal field theory is investigated using the ADS/CFT correspondence
and a novel method based on nilpotent weights. Using this device we add
ghost fermions and point to a BRST invariance of the theory. \vspace{5mm}%
\end{abstract}

\section{Introduction}

Much work has been done in the last few years based on the AdS/CFT
correspondence with the aim of understanding conformal field
theories \cite {ADS}. Within this framework, the correlation
functions of operators on the boundary of Anti de Sitter space
are determined in terms of appropriate bulk propagators. While
the form of the two and three point functions within CFT are
fixed by conformal invariance, it is interesting to find actions
in the bulk which result in the desired boundary green functions.
In particular it is interesting to discover which actions give
rise to logarithmic conformal field theories (LCFTs) which leads
to AdS/LCFT correspondence. Two points should be clarified, first
what is meant by ordinary AdS/CFT correspondence and second what
is an LCFT, and how does it fit into the correspondence.

The conjecture states that a correspondence between theories
defined on AdS$_{d+1}$ and CFT$_{d}$ can be found. Suppose that a
classical theory is defined on the AdS$_{d+1}$ via the action
$S[\Phi]$. On the boundary of this space the field is constrained
to take certain boundary value ${\Phi|}_{\partial AdS} =\Phi_0$.
With this constraint, one can calculate the partition function
\begin{eqnarray}
Z[\Phi_0]={e^{-S_{Cl}[\Phi]}}|_{\Phi_{\partial AdS}=\Phi_0}.
\end{eqnarray}

On the other hand in the CFT$_{d}$ space, there exist operators
like $\hat{O}$ which belong to some conformal tower. Now the
correspondence states that the partition function calculated in
AdS is the generating function of the theory in CFT with $\Phi_0$
being the source, that is

\begin{equation}\label{AdS/CFT}
Z[\Phi_0]=\left\langle e^{\int\hat{O}\Phi_0} \right\rangle.
\end{equation}

How do logarithmic conformal field theories fit into this
picture? A decade after the seminal paper by Belavin, Polyakov and
Zamolodchikov \cite {BPZ} on the determining role of conformal
invariance on the structure of two dimensional quantum field
theories, Gurarie \cite{Gur} pointed to the existence of
(LCFT's). Correlation functions in an LCFT may have logarithmic
as well as power dependence \cite{Kogan}. Such logarithmic terms
were ruled out earlier due to requirements such as unitarity or
non existence of null states. The literature on LCFT is already
very long, for a survey see some of the recent papers on LCFT for
example \cite{FlohrNew}, \cite{Lewis}.

The bulk actions defined on AdS$_{3}$ which give rise to
logarithmic operators on the boundary where first discussed in
\cite{Khorrami,KoganAds} and have consequently been discussed by
a number of authors \cite{Lewis,many}. More recently a connection with
world sheet supersymmetry has been discussed in
\cite{KoganPolyakov}.

In an LCFT, degenerate operators exist which form a Jordan cell
under conformal transformations. In the simplest case, one has a
pair $\hat{A}$ and $\hat{B}$ transforming as
\begin{eqnarray}
\hat{A}(\lambda z)&=&\lambda^{-\Delta}\hat{A}(z) ,  \nonumber \\
\hat{B}(\lambda z)&=&\lambda^{-\Delta}[\hat{B}(z)-\hat{A}(z)\ln
\lambda] .
\end{eqnarray}

In references \cite{MRS,MRSAlgeb} we modified the method developed
by Flohr \cite{Flohr} in which nilpotent variables were
introduced. We defined a 'superfield' :
\begin{equation} \label{OF}
\hat{O}(z,\eta)= \hat{A}(z) + \hat{\bar{\zeta}}(z) \eta
+\bar{\eta} \hat{\zeta}(z) + \bar{\eta} \eta \hat{B}(z),
\end{equation}
using different components of a logarithmic pair and adding
fermionic fields. The word 'superfield' here does not mean that a
supersymmetric invariance exists, rather it is a convenient tool.

Now we observe that $\hat{O} (z,\eta )$ has the following
transformation law under scaling
\begin{eqnarray}  \label{TRLaw}
\hat{O}(\lambda z,\eta)=\lambda^{-(\Delta+\bar{\eta}\eta)}
\hat{O}(z,\eta) .
\end{eqnarray}

To find out what this scaling law means, one should expand both
sides of equation (5) in terms of $\eta$ and $\bar{\eta}$. Doing
this and comparing the two sides of equation (5), it is found that
$\hat{A}(z)$ and $\hat{B}(z)$ transform as equation (3) and
$\zeta$ and $\bar{\zeta}$ are ordinary fields of dimension
$\Delta$. The appearance of such fields has been proposed by
Kausch \cite{Kausch}, within the $c=-2$ theory. As discussed in
\cite{MRS,MRSAlgeb} using this structure one can derive most of
the properties of LCFTs.

This paper is organized as follows: in section 2 the
correspondence is explained explicitly and the two point
correlation functions of different fields of CFT are derived. In
section 3 we state that there should be a BRST invariance in the
theory due to existence ghost fields and and find the BRST
transformation in both AdS and LCFT spaces and finally show the
compatibility of the correlation functions derived in section 2
with the BRST symmetry.

\section{AdS/LCFT Correspondence and Correlation Functions}

 To begin, one should propose an action on AdS. As we will have an
operator just like the one in equation ($\!\!$~\ref{OF}) in LCFT
part of the theory, there should be a corresponding field in AdS,
$\Phi(\eta)$, which can be expanded as

\begin{equation}\label{PHI}
\Phi(x,\eta)=C(x)+\bar{\eta}\alpha(x)+\bar{\alpha}(x)\eta+
\bar{\eta}\eta D(x),
\end{equation}
where $x$ is $(d+1)-$dimensional with components $x^0,\cdots,x^d$.
Of course $d=2$ is the case we are most interested in, any how the
result can be applied to any dimension and so we will consider
the general case in our calculations.

Let us then consider the action:
\begin{equation}
S=-\frac 12\int d^{d+1}x\int d\bar{\eta} d\eta [(\nabla \Phi
(x,\eta )).(\nabla \Phi (x,-\eta ))+ m^2(\eta )\Phi(x,\eta )\Phi
(x,-\eta )]. \label{Action}
\end{equation}

This action seems to be the simplest non-trivial action for the
field $\Phi (\eta ).$ To write it explicitly in terms of the four
components of the the field, one should expand equation (7) in
powers of $\bar{\eta}$ and $\eta$ using equation (6). Integrating
over $\bar{\eta}$ and $\eta$ one finds
\begin{equation}  \label{Action_a}
S=-\frac{1}{2}\int d^{d+1} x [2(\nabla C).(\nabla D)+ 2{m^2}_1CD
+{m^2}_2 C^2 + 2(\nabla\bar{\alpha}).(\nabla\alpha)+2{m^2}_1
\bar{\alpha}\alpha].
\end{equation}
To derive expression (8) we have assumed $m^2(\eta)={m^2}_1
+{m^2}_2 \bar{\eta} \eta$. Note that the bosonic part of this
action is the same as the one proposed by
\cite{Khorrami,KoganAds} with ${m^2}_1=\Delta(\Delta-d)$ and ${
m^2}_2=2\Delta-d$. In our theory with proper scaling of the
fields, one can recover these relations.

The equation of motion for the field $\Phi$ is
\begin{equation}  \label{Eq.Motion}
({\nabla}^2 - m^2(\eta))\Phi(x,\eta)=0.
\end{equation}
The Drichlet Green function for this system satisfies the equation
\begin{equation}  \label{eqGreen}
({\nabla}^2 - m^2(\eta))G(x,y,\eta)=\delta(x,y)
\end{equation}
together with the boundary condition
\begin{equation}  \label{Boundary}
G(x,y,\eta)|_{x\in \:\partial_{AdS}} =0.
\end{equation}

With this Green function, the Drichlet problem for $\Phi$ in AdS
can be solved readily. However, near the boundary of AdS, that is
$x^d \simeq 0$,  the metric diverges so the problem should be
studied more carefully. One can first solve the problem for the
boundary at $x^d=\varepsilon$ and then let $\varepsilon$ tend to
zero. With properly redefined scaled fields at the boundary one
can avoid the singularities in the theory. So we first take the
boundary at $ x^d=\varepsilon$ and find the Green function
\cite{Muck}
\begin{equation}\label{Green}
G(x,y,\eta)|_{y^d=\varepsilon}=-a(\eta)\varepsilon^
{\Delta+\bar{\eta}\eta-d}
\left(\frac{x^d}{(x^d)^2+|\mathbf{x-y}|^2}\right)^
{\Delta+\bar{\eta}\eta}.
\end{equation}
where the bold face letters are $d$-dimensional and live on the
boundary. The field in the bulk is related to the boundary fields
by
\begin{equation}  \label{b to b}
\Phi(x,\eta)=2a(\eta)(\Delta+\bar{\eta}\eta)\varepsilon^{\Delta+\bar{\eta}
\eta-d} \int_{y^d=\varepsilon} d^d y\:
\Phi(\mathbf{y},\varepsilon,\eta)
\left(\frac{x^d}{(x^d)^2+|\mathbf{x-y}|^2}\right)^
{\Delta+\bar{\eta}\eta}
\end{equation}
with $a=\frac{\Gamma(\Delta+\bar{\eta}\eta)}{2\pi^{d/2}
\Gamma(\alpha+1)}$ and $\alpha=\Delta+\bar{\eta}\eta-d/2$. To
compute Gamma functions in whose argument appears
$\bar{\eta}\eta$, one should make a Taylor expansion for the
function, that is: \begin{equation}
\Gamma(a+\bar{\eta}\eta)=\Gamma(a)+\bar{\eta}\eta\Gamma^{\prime}(a).
\end{equation} There is no higher terms in this Taylor expansion
because $(\bar{\eta} \eta)^2=0$.

Now defining
$\Phi_b(\mathbf{x},\eta)=\lim_{\varepsilon\rightarrow0} (\Delta+
\bar{\eta}\eta)\varepsilon^{\Delta+\bar{\eta}\eta-d}\Phi(\mathbf{x}
,\varepsilon,\eta)$, we have
\begin{equation}
\Phi(x,\eta)=\int d^d \mathbf{y}
\left(\frac{x^d}{(x^d)^2+|\mathbf{x-y}|^2} \right)^
{\Delta+\bar{\eta}\eta} \Phi_b(\mathbf{y},\eta).
\end{equation}

Using the solution derived, one should compute the classical action. First
note that the action can be written as (using the equation of motion and
integrating by parts)
\begin{equation}  \label{j}
S_{cl.}=\frac{1}{2} \lim_{\varepsilon\rightarrow 0}
\varepsilon^{1-d} \int d \bar{\eta}d \eta \int d^d \mathbf{y}
\left[\Phi(\mathbf{y},\varepsilon,\eta) \frac{\partial
\Phi(\mathbf{y},\varepsilon,-\eta)}{\partial x^d}\right]
\end{equation}
putting the solution (13) into equation (14), the classical
action becomes
\begin{equation}  \label{dd}
S_{cl.}(\Phi_b)=\frac{1}{2}\int d \bar{\eta}d \eta \int d^d
\mathbf{x} d^d \mathbf{y} \frac{a(\eta)
\Phi_b(\mathbf{x},\eta)\Phi_b(\mathbf{y},-\eta)}{|
\mathbf{x-y}|^{2\Delta+2\bar{\eta}\eta}}.
\end{equation}

The next step is to derive correlation functions of the operator
fields on the boundary by using AdS/CFT correspondence, i.e.
equation (2). In our language the operator $\hat{O}$ has an $\eta
$ dependence, in addition to its usual coordinate dependence. It
lives in the LCFT space and can be expanded as
\begin{equation}
\hat{O}(\mathbf{x},\eta)=\hat{A}(\mathbf{x})+\bar{\eta}\hat{\zeta}(\mathbf{x})+
\hat{\bar{\zeta}}(\mathbf{x})\eta +\bar{\eta}\eta
\hat{B}(\mathbf{x}),
\end{equation}
so the AdS/CFT correspondence becomes
\begin{equation}
\left\langle exp\left( \int d\bar{\eta}d\eta \int
d^{d}\mathbf{x}\hat{O}(\mathbf{x},\eta )\Phi _{b}
(\mathbf{x},\eta )\right) \right\rangle =e^{S_{cl}(\Phi _{b})}.
\label{AdS/LCFT}
\end{equation}
Expanding both sides of equation (19) in powers of $\Phi _{b}$
and integrating over $\eta $'s, the two-point function of
different components of $\hat{O}(\mathbf{x},\eta )$ can be found
\begin{eqnarray}
\langle \hat{A}(\mathbf{x})\hat{A}(\mathbf{y})\rangle &=&0
\label{corr} \\ \ \langle \hat{A}(\mathbf{x})\hat{B}(\mathbf{y})
\rangle &=&\frac{a_{1}} {(\mathbf{x}-\mathbf{y})^{2\Delta }} \\
\langle \hat{B}(\mathbf{x})\hat{B}(\mathbf{y})\rangle  &=&
\frac{1}{(\mathbf{x}-\mathbf{y})^{2\Delta
}}(a_{2}-2a_{1}\log (\mathbf{x}-\mathbf{y})) \\ \langle
\hat{\bar{\zeta}}(\mathbf{x})\hat{\zeta}(\mathbf{y})\rangle
&=&\frac{-a_{1}}{(\mathbf{x}-\mathbf{y})^{2\Delta }}
\end{eqnarray}
with all other correlation functions being zero. These
correlation functions can be obtained in another way. Knowing the
behaviour of the fields under conformal transformations, the form
of two-point functions are determined. The scaling law is given
by equation (5). Using this scaling law, most of the correlation
functions derived here are fulfilled, however it does not lead to
vanishing correlation functions of $ \langle
\hat{A}(\mathbf{x})\hat{\zeta}(\mathbf{y})\rangle $ and $\langle
\hat{B}(\mathbf{x})\hat{\zeta}(\mathbf{y}) \rangle $. These
correlation functions are found to be:
\begin{eqnarray} \langle
\hat{A}(\mathbf{x})\hat{\zeta}(\mathbf{y})\rangle &=& \frac{b_1}
{\mathbf{(x-y)}^{2\Delta}}
\\ \langle \hat{B}(\mathbf{x})\hat{\zeta}(\mathbf{y})\rangle &=&
\frac{1}{\mathbf{(x-y)}^{2\Delta}} \left( b_2 -2b_1
\log\mathbf{(x-y)} \right).
\end{eqnarray}

Of course, assuming $b_1=b_2=0$ one finds the forms derived above.
However, the vanishing value of such correlators comes from some
other properties of the theory. What forces these constants to
vanish is the fact that the total fermion number is odd
\footnote{this observation is due to M. Flohr}. One way of seeing
this is to look at the OPE as given in \cite{MRSAlgeb}. The OPE
of two $\hat{O}$-fields has been proposed to be:
\begin{equation}\label{OPE}
\hat{O}(z)\hat{O}(0)\sim z^{\bar{\eta_1}\eta_2+\bar{\eta_2}\eta_1}
\frac{\hat{\Phi}_0(\eta_3)}{z^{2(\Delta+\bar{\eta_3}\eta_3)}}
\end{equation}
where $\eta_3=\eta_1+\eta_2$ and $\hat{\Phi}_0$ is
the identity multiplet:
\begin{equation}\label{Identity}
\hat{\Phi}_0(\eta)=\hat{\Omega} +
\bar{\eta}\hat{\xi}+\hat{\bar{\xi}}\eta+\bar{\eta}\eta\hat{\omega}
\end{equation}
with the property
\begin{equation}
\langle\hat{\Phi}_0(\eta)\rangle=\bar{\eta}\eta.
\end{equation}
Note that the ordinary identity operator is $\hat{\Omega}$ which
has the unusual property that $\langle\hat{\Omega}\rangle=0$, but
its logarithmic partner, $\hat{\omega}$, has nonvanishing norm.

Calculating the expectational value of both sides of equation
(26), the correlation functions of different fields inside
$\hat{O}$ are found which leads to vanishing correlators $ \langle
\hat{A}(\mathbf{x})\hat{\zeta}(\mathbf{y})\rangle $ and $\langle
\hat{B}(\mathbf{x})\hat{\zeta}(\mathbf{y}) \rangle $, just the
same result as derived by ADS/LCFT correspondence.
\section{BRST symmetry of the theory}
The existence of some ghost fields in the action of the theory
considered so far, is reminiscent of BRST symmetry. A few remarks
clarifying the word "ghost" may be in order here. The action
defined by equation (8) has two types of fields in it. The usual
scalars as used by other authors $C$ and $D$, and fermionic fields
$\alpha$ which have been added due to the superfield structure. We
have called the fermionic fields "ghosts" because they are scalar
fermions and also because they participate in the BRST symmetry
as we shall see below.

As the action is quadratic, the partition function can be
calculated explicitly:

\begin{equation}
Z=\int{\rm\bf D}C(x){\rm\bf D}D(x){\bf D}\bar{\alpha}(x){\bf
D}\alpha(x)e^{-S[C,D,\bar{\alpha},\alpha]}.
\end{equation}

The bosonic and fermionic parts of the action are decoupled and
integration over each of them can be performed independently. For
bosonic part one has
\begin{equation}
Z_{b} = \int\prod_pdC_{p}\int\prod_pdD_p\exp \left\{-\frac{1}{2}
(\begin{array}{ll} C_p & D_p
\end{array})
G(p) \left(\begin{array}{l}
C_p \\
D_p
\end{array}\right)\right\}
\end{equation}
where a Fourier transform has been done and
\begin{equation}
G(p)=\left(\begin{array}{cc}
  p^2+{m^2}_1 & 0 \\
  {m^2}_2 & p^2+{m^2}_1
\end{array}\right).
\end{equation}

These integrals are simple Gaussian ones. So, apart from some
unimportant numbers, this partition function becomes
\begin{equation}
Z_b=\prod_p\left[\det\left(\begin{array}{cc}
  p^2+{m^2}_1 & 0 \\
  {m^2}_2 & p^2+{m^2}_1
\end{array}\right)\right]^{-1/2}=\prod_p(p^2+{m^2}_1)^{-1}.
\end{equation}

For the fermionic part the the same steps can be done and the
result is
\begin{equation}
Z_f=\prod_p(p^2+{m^2}_1).
\end{equation}
Now it is easily seen that the total partition function is merely
a number, independent of the parameters of the theory and this is
the signature of BRST symmetry. Before proceeding, it is worth
mentioning that this symmetry will be induced onto LCFT part of
the correspondence and the correlation functions in that space
will also be invariant under proper transformations.

In BRST transformation, the fermionic and bosonic fields should
transform into each other. In our case this this can be done using
$\eta$ and $\bar{\eta}$. Also one needs an infinitesimal
anticommuting parameters. Now let $\epsilon_1$ and $\epsilon_2$
be infinitesimal anticommuting parameters and consider the
following infinitesimal transformation of the the field $\Phi$
\begin{equation}\label{BRST Tr.}
\delta\Phi(\mathbf{x},\eta)=(\bar{\epsilon}\eta +
\bar{\eta}\epsilon) \Phi(\mathbf{x},\eta).
\end{equation}
It can be easily seen that this transformation leaves the action
invariant, because in the action the only terms which exist are
in the form of $\Phi(\eta)\Phi(-\eta)$ and under such a
transformation this term will become
\begin{equation}\label{delta phi2}
\delta(\Phi(\eta)\Phi(-\eta))=\Phi(\eta)(-\bar{\epsilon}\eta -
\bar{\eta}\epsilon) \Phi(-\eta)+(\bar{\epsilon}\eta +
\bar{\eta}\epsilon) \Phi(\eta)\Phi(-\eta)
\end{equation}
which is identically zero. We can therefore interpret this
transformation as the action of two charges $Q$ and $\bar{Q}$. The
explicit action of $Q$ for each component of $\Phi$ is :
\begin{eqnarray}
Q C&=&0 \\
Q \alpha&=&0 \\
Q \bar{\alpha}&=&C\\
Q D&=&-\alpha
\end{eqnarray}
As expected the bosonic and fermionic fields are transformed into
each other, and the square of $Q$ vanishes. The action of
$\bar{Q}$ is similar except that it vanishes on $\bar{\alpha}$
and not on $\alpha$.

To see how this symmetry is induced onto the LCFT part of the
theory one should first find the proper transformation. Going back
to equation (19) and using the symmetry obtained for the classical
action one finds
\begin{equation}
\exp(S_{Cl}[\Phi])=
\left\langle\exp \left(\int\hat{O}(\Phi+ \delta\Phi)\right)
\right\rangle
\end{equation}
As the transformation of $\Phi$ is $(\bar{\epsilon}\eta +
\bar{\eta}\epsilon)\Phi$ the integrand on the right hand side of
equation (40) is just $\hat{O}\Phi+(\bar{\epsilon}\eta +
\bar{\eta}\epsilon)\hat{O}\Phi$ which can be regarded as
$(\hat{O}+\delta\hat{O})\Phi$ with
$\delta\hat{O}=(\bar{\epsilon}\eta + \bar{\eta}\epsilon)\hat{O}$.
So equation (40) can be rewritten as
\begin{equation}
\left\langle\exp\left(\int\hat{O}\Phi\right)\right\rangle=
\left\langle\exp\left(\int(\hat{O}+\delta\hat{O})
\Phi\right)\right\rangle.
\end{equation}
This shows that the correlation functions of the $\hat{O}$ field
are invariant under the BRST transformation, that is
$\delta\langle \hat{O}_1\hat{O}_2\cdots\hat{O}_n\rangle=0$ if the
BRST transformation is taken to be
\begin{equation}
\delta \hat{O}=(\bar{\epsilon}\eta + \bar{\eta}\epsilon)\hat{O}.
\end{equation}
Again one can rewrite this transformation in terms of the
components of $\hat{O}$ and the result is just the same as
equations (36)-(39).

This invariance can be tested using two point correlation
functions derived in previous section. These correlation functions
are easily found to be invariant under the transformation law
(42), as an example
\begin{equation}\label{example}
Q \langle B \bar{\zeta}\rangle= \langle (Q B) \bar{\zeta}\rangle+
\langle B (Q \bar{\zeta}) \rangle= \langle\bar{\zeta}\zeta\rangle
- \langle BA\rangle=0.
\end{equation}

{\bf \Large Conclusion}

Logarithmic conformal field theory has been investigated using a
novel method based on nilpotent weights. This method also allows
investigation of LCFTs within the AdS/CFT correspondence. Although
the emergence of LCFTs through this correspondence had been noted
before, but the present method is much easier to implement. Also
the transparency of the present method allows inclusion of
fermionic fields, thus pointing to a novel symmetry both in the
Ads and in LCFT.

{\bf \Large Acknowledgements}

We wish to thank Michael Flohr for constructive comments.

\end{document}